\documentclass[aps,twocolumn,pra,groupedaddress,superscriptaddress,amsmath,amssymb,showpacs,noeprint]{revtex4-2}

\usepackage{graphicx}
\usepackage{dcolumn}
\usepackage{bm}
\usepackage{hyperref}
\usepackage{mathtools, cuted}
\usepackage{verbatim}
\usepackage[mathlines]{lineno}
\usepackage{physics}
\usepackage{xfrac}

\usepackage{soul}
\usepackage{color}
\usepackage{csquotes}
\usepackage{comment}

\newcommand{\prt}[1]{\left(#1\right)}
\newcommand{\prtq}[1]{\left[#1\right]}
\newcommand{\prtg}[1]{\left\{#1\right\} }
\newcommand{\ad}{a^\dagger}
\newcommand{\rth}{\textup{rth}}

\newcommand{\bath}{\textup{bath}}
\newcommand{\prtb}[1]{\Bigg(#1\Bigg)}
\newcommand{\rot}{\rho^{\textup{th}}}
\newcommand{\rotr}{\rho^{\textup{rth}}}
\newcommand{\N}{\mathcal{N}}

\newenvironment{equations}
    {
    \begin{equation}
    \begin{aligned}
    }
    {
    \end{aligned}
    \end{equation}\ignorespacesafterend
    }

\begin{document}

\title{Simple scheme for extracting work with a single bath}

\author{Nicol\`o Piccione}
\email{nicolo.piccione@univ-fcomte.fr}

\affiliation{Institut UTINAM, CNRS UMR 6213, Universit\'{e} Bourgogne Franche-Comt\'{e}, Observatoire des Sciences de l'Univers THETA, 41 bis avenue de l'Observatoire, F-25010 Besan\c{c}on, France}

\author{Benedetto Militello}
\affiliation{Universit\`a degli Studi di Palermo, Dipartimento di Fisica e Chimica - Emilio segr\`e, via Archirafi 36, I-90123 Palermo, Italy}
\affiliation{Istituto Nazionale di Fisica Nucleare, Sezione di Catania, via Santa Sofia 64, I-95123 Catania, Italy}

\author{Anna Napoli}
\affiliation{Universit\`a degli Studi di Palermo, Dipartimento di Fisica e Chimica - Emilio segr\`e, via Archirafi 36, I-90123 Palermo, Italy}
\affiliation{Istituto Nazionale di Fisica Nucleare, Sezione di Catania, via Santa Sofia 64, I-95123 Catania, Italy}

\author{Bruno Bellomo}
\affiliation{Institut UTINAM, CNRS UMR 6213, Universit\'{e} Bourgogne Franche-Comt\'{e}, Observatoire des Sciences de l'Univers THETA, 41 bis avenue de l'Observatoire, F-25010 Besan\c{c}on, France}

\begin{abstract}

We propose a simple protocol exploiting the thermalization of a \emph{storage} bipartite system $S$ to extract work from a \emph{resource} system $R$.
The protocol is based on a recent work definition involving only a single bath.
A general description of the protocol is provided without specifying the characteristics of $S$.
We quantify both the extracted work and the ideal efficiency of the process also giving  maximum bounds for them.
Then, we apply the protocol to two cases: two interacting qubits and the Rabi model.
In both cases, for very strong couplings, an extraction of work comparable with the bare energies of the subsystems of $S$ is obtained and its peak is reached for finite values of the bath temperature, $T$.
We finally show, in the Rabi model at $T=0$, how to transfer the work stored in $S$ to an external device, permitting thus a cyclic implementation of the whole work-extraction protocol.
Our proposal makes use of simple operations not needing fine control.

\end{abstract}

\maketitle

\section{\label{sec:Introduction}Introduction}

In recent years, interest in quantum thermodynamics has been growing (for a review, see \cite{Vinjanampathy2016}).
One of the most intriguing problems concerns the realization of thermodynamic processes at a quantum level \cite{Quan2007,Quan2006,Kim2011,Humphrey2002,Leggio2015,James2016,Hewgill2018}.
Other topics range from typicality \cite{Tipicality1,Tipicality2} to maximum entropy production principle \cite{Martyushev2006,Beretta2009,Beretta2014,Militello2018}.
Recently, many results have been obtained inside the theoretical framework of the thermodynamic resource theory (TRT).
Among them, we cite the thermomajorization requirement \cite{Horodecki2013} and the generalized second laws \cite{Brandao2015,Cwiklinski2015,Lostaglio2015}.
However, these theorems are derived assuming as admissible also very complex thermodynamic processes so that the experimental realization of these protocols could be unfeasible.

Among all the possible quantum processes, work-extraction protocols play a relevant role \cite{Horodecki2013,Skrzypczyk2014}.
However, most of them are not easy to realize experimentally. For this
reason, various efforts have been made to understand how to
design realizable thermodynamic protocols \cite{Lostaglio2018,Perry2018}.
Nevertheless, most proposals require fine control of the system for an
experimental realization. For example, in a process composed
of many steps, it could be required to turn on and off a specific
interaction for an amount of time specific to each step.

In this paper, we conceive a work-extraction protocol exploiting a single bath and making use of simple operations
which should be easily implementable without need for fine operations. Indeed, we propose to extract work from a resource system R to a bipartite quantum system S exploiting simple operations such as a thermalization process and turning on and off the interaction between the two subsystems of S~\cite{Piccione2018}.
We show that this thermalization protocol gives rise to
a quite efficient single-shot work extraction.
To quantify the work extracted, we make use of a work quantifier recently introduced in the context of TRT \cite{WorkDef}.
In order to make the procedure cyclic, we show in one of the considered models how to exploit the result of the thermalization protocol to charge an external device playing the role of a quantum
battery through a quite simple transfer protocol. We stress
that the various parts of the global protocol do not need
fine control. For example, the interaction between the two
subsystems of $S$ does not need to last for a precise amount
of time but only enough to let system $S$ thermalize, while
the procedures of switching on and off have to be just rapid
enough to leave unaltered the state of the system.

To better appreciate the potentialities of our protocol we apply it to two different physical scenarios (a two-qubit system and a spin-boson system) described by different models.
The first model can describe the interaction of two spins in an Ising chain \cite{Kadowaki1998,Dawson2004,Decordi2017}, while the second is described by the ubiquitous Rabi Hamiltonian.
The latter model is very effective for example in cavity QED \cite{Haroche_book} and in circuit QED \cite{QRM_Experiment2}.
In the past decades this model has been mainly treated under suitable approximations such as the rotating wave approximation \cite{Haroche_book,Petruccione_book} and the Bloch-Siegert approximation \cite{Bloch-Siegert}, which hold when the interaction is weak.
Recently, an analytical complete solution has been found \cite{Braak,QRM_Bogoliubov} (see also \cite{Quantum_Rabi_Review} for a review)
and a lot of attention has been devoted, both theoretically and experimentally \cite{USC_definition,Deep_Strong_Coupling_definition,QRM_Experiment,QRM_Experiment2,QRM_Experiment3,Generalized_Rotating-Wave_Approximation,Braak,QRM_Bogoliubov,Quantum_Rabi_Review}, to the study of the Rabi Hamiltonian beyond the weak coupling regime, also in view of the recent remarkable experimental realizations of physical situations characterized by high values of the interaction strength \cite{QRM_Experiment,QRM_Experiment2,QRM_Experiment3}.

The paper is organized as follows.
In Sec.~\ref{sec: Work extraction protocol}, we describe the thermalization protocol for an arbitrary bipartite system.
In Sec.~\ref{sec: Two interacting qubits}, we describe a possible realization of our protocol in a system consisting of two qubits, while in Sec.~\ref{sec: Rabi model} we consider a spin-boson system the interaction of which is described by the Rabi model.
We also discuss the possibility to transfer the extracted work to another physical system, effectively charging a battery.
Finally, in Sec.~\ref{sec:Conclusions} we provide some conclusive remarks on our results.
Some details of our analysis can be found in the Appendices.

\section{\label{sec: Work extraction protocol} Work-extraction protocol}

\subsection{\label{subsec: Work quantifier}Work quantifier}

In this paper, we choose a work quantifier within the framework of TRT among those described in Ref.~\cite{WorkDef}, which is strictly connected to the von Neumann free energy (see Appendix~\ref{sec: Principles of thermodynamic resource theory} for a brief overview of TRT). Our specific choice of the work quantifier is motivated by the fact that, differently from other quantifiers, it can be used even if the resource $R$ (which can be classical or quantum) and the storage $S$ are correlated at the end of the process~\cite{WorkDef}.
The quantifiers treated in Ref.~\cite{WorkDef} are analyzed by considering the set of possible processes described by TRT for a fixed environmental temperature and, moreover, they have to respect some axioms built in such a way that the second law of thermodynamics is automatically satisfied.
These quantifiers assess how much the \enquote{usefulness} of a system has changed after an operation, when dealing with an environment at a fixed temperature.

In what follows,
the work stored in system $S$ during the process is given by:
\begin{equation}
\label{eq: Eisert definition}
    W=\Delta F\prt{\rho_S',H_S'}-\Delta F(\rho_S,H_S),
\end{equation}
where
\begin{equation}
\label{eq: definition delta F}
\begin{aligned}
\Delta F(\rho,H)&=F(\rho,H)-F\left(\rho^{\textup{th}},H \right), \\
F(\rho,H)&=\Tr \prtg{H\rho} -k_B TS(\rho), \\
S(\rho) &=-\Tr\prtg{\rho \ln (\rho)}.
\end{aligned}
\end{equation}
Here, $F(\rho,H)$ is the free energy of the state $\rho$ when the system is governed by the Hamiltonian $H$, $\rho^{\textup{th}}$ is the thermal state of the system, corresponding to the Hamiltonian $H$ at temperature $T$ equal to the temperature of the thermal bath which is used in the process, $k_B$ is Boltzmann constant, $S(\rho)$ is the von Neumann entropy of the state $\rho$ and $H_S$ is the Hamiltonian of system $S$.
The quantities marked with an apex are related to the end of the process, while those not marked are related to the start of the process.
We remark that this work definition quantifies how much the \enquote{usefulness} amount of a system, given by $\Delta F(\rho,H)$, changes after a permitted operation.
If $H_S'=H_S$, Eq.~\eqref{eq: Eisert definition} simplifies to:
\begin{equation}
\label{eq: work definition no Hamiltonian change}
    W=F\prt{\rho_S',H_S}-F\prt{\rho_S,H_S}.
\end{equation}
Then, at zero temperature and for a nonchanging Hamiltonian, the chosen definition of work coincides with the average energy difference of system $S$ between the start and the end of the protocol, i.e., with the intuitive definition of work done on a system.
Some comments on the possible links between this work quantifier and other thermodynamic quantities can be found in Appendix~\ref{sec: Work definition, heat and entropy production}

\subsection{\label{subsec: The phases of the protocol}The thermalization protocol}

\begin{figure}
    \centering
    \includegraphics[width=0.47\textwidth]{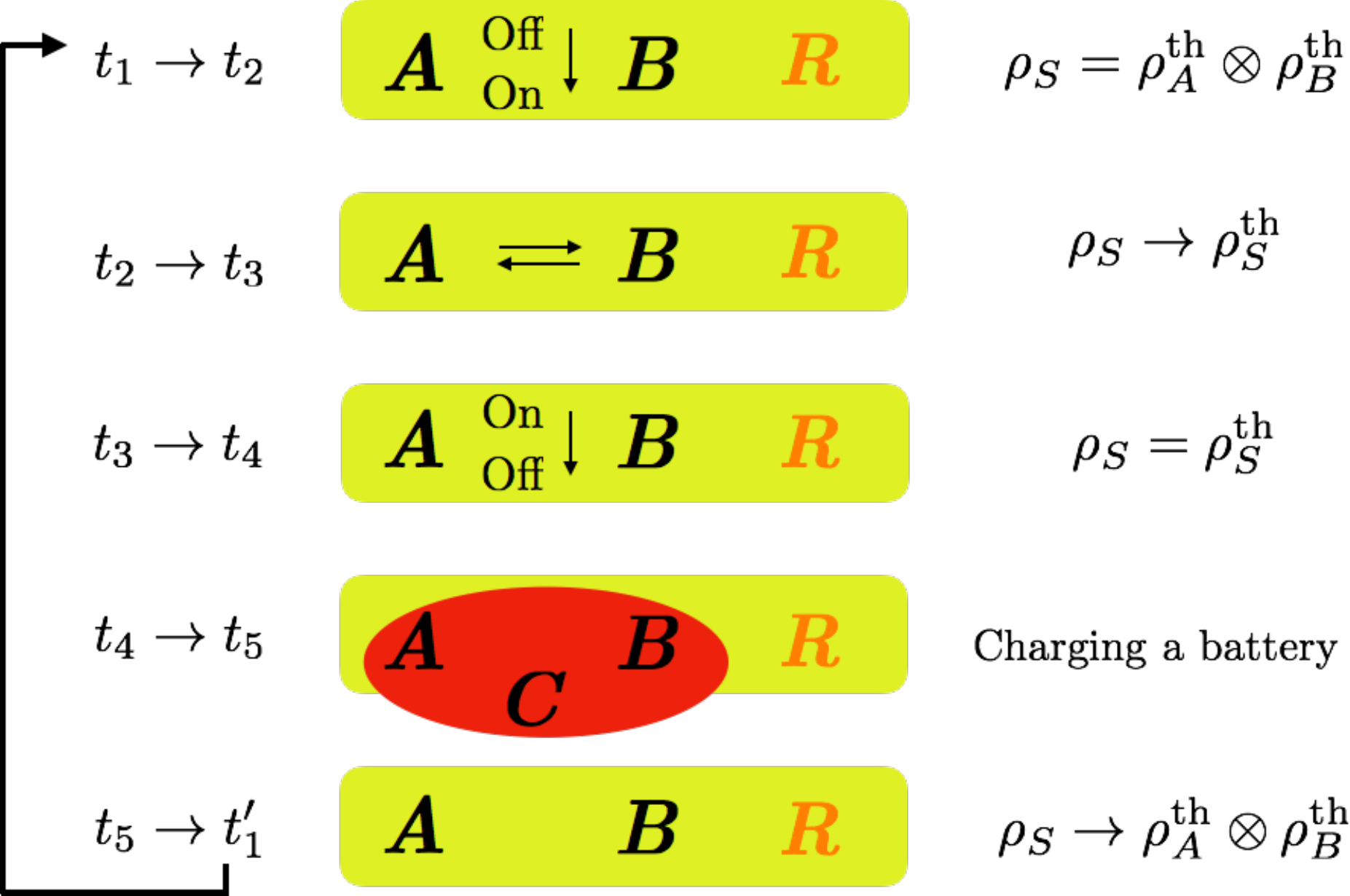}
    \caption{Schematic illustration of the phases of the thermalization protocol in the presence of a thermal bath (yellow box). At the start ($t_1$) the subsystems do not interact. Then, the interaction is turned on and they thermalize together (from $t_2$ to $t_3$). Once they have thermalized their interaction can be turned off and the process of work extraction is completed ($t_4$). If one wants to transfer the extracted work to iterate the process one does so in the time interval $\delta_T t$ and, after having finished the process ($t_5$), the experimental setup might have to be reinitialized, from $t_5$ to $t_1'$, where $t_1'$ plays the role of the initial time of the new cycle. }
    \label{fig:Grafico_protocollo}
\end{figure}

Here, we describe the thermalization protocol in the case $S$ is an arbitrary bipartite system composed of two subsystems $A$ and $B$.
The protocol  can be divided into different phases (see Fig.~\ref{fig:Grafico_protocollo}).
For each phase, we compute the free energy using the notation $F_i=F(\rho_S (t_i), H (t_i))$.
Notice that, the presence of the environment is necessary during the thermalization from $t_2$ to $t_3$. During the other phases it would be enough to assume  that  the environment is at disposal if needed. However, in  the following analysis we always refer to a realistic situation where the environment and system $S$ are always interacting.
In this case, we must assume that the interaction is so weak that, overall, whenever we have to take into account the evolution given by the interaction of $S+R$ with the bath, the total energy $\ev{H_S + H_R+H_\bath}$, where $H_R$ and $H_\bath$ are, respectively, the Hamiltonians of the resource $R$ and of the bath, is a conserved quantity and results from TRT can be applied.

At the start of the protocol ($t=t_1$) $A$ and $B$ are non-interacting, non correlated and spatially separated \footnote{The spatial separation is not strictly required, but we think it facilitates the experimental implementation because it is, usually, easy to turn on and off the interaction between two systems by just joining and separating them spatially.}.
They are both in contact with the same thermal bath at temperature $T$ so that:
\begin{equation}
\label{eq: status t1}
\begin{aligned}
H(t_1)&=H_A+H_B=H_0, \quad \rho_{S}(t_1)=\rot_A\otimes \rot_B,\\
F_1 &=F\prt{\rho_A^{\textup{th}},H_A}+F\prt{\rho_B^{\textup{th}},H_B},
\end{aligned}
\end{equation}
where $H_{A(B)}$ is the free Hamiltonian of $A(B)$, $\rho_{A(B)}^{\textup{th}}=e^{-\beta H_{A(B)}}/ \Tr \prtg{\mathrm{e}^{-\beta H_{A(B)}}}$ is the local thermal state and $\beta=1/(k_B T)$.
Then, the interaction between $A$ and $B$ is turned on by exploiting $R$, during a time interval $\delta_1 t$ from $t_1$ to $t_2=t_1+\delta_1 t$.
We suppose that the state of system $S$ does not change during this time.
This can be achieved if the turning on of the interaction is doable in a time interval much smaller than the typical evolution time of system $S$, during the switching on procedure, coupled to the bath. Defining  $\ev{H}_{t_n}=\Tr \prtg{\rho_S (t_n) H}$,
we have:
\begin{equation}\label{eq: status t2}
\begin{aligned}
H(t_2)&=H_0+H_I, \quad \rho_{S}(t_2)=\rho_{S}(t_1),\\
F_2 &=F_1+\ev{H_I}_{t_2},
\end{aligned}
\end{equation}
$H_I$ being  the interaction Hamiltonian between  $A$ and $B$.

From time $t_2$ to time $t_3$ ($t_3-t_2 \gg \tau_r$, where $\tau_r$ is the typical evolution time of system $S$ in this phase), $A$ and $B$ thermalize as a whole so that at $t_3$:
\begin{equations}
\label{eq: status t3}
H(t_3)&=H(t_2), \quad\rho_{S}(t_3)=\rho_{S}^{\textup{th}},\\
F_3 &=F\prt{\rho_{S}^{\textup{th}},H_0}+\ev{H_I}_{t_3},
\end{equations}
where $ \rho^{\textup{th}}_S=e^{-\beta (H_0+H_I)}/\Tr \prtg{e^{-\beta  (H_0+H_I)}}$ is the global thermal state.
Finally, from time $t_3$ to time $t_4=t_3 + \delta_2 t$ we use again system $R$ to turn off the interaction term between subsystems $A$ and $B$ by spatially separating them \cite{Note1} and supposing that the state of $S$ remains unaltered.
The situation at time $t_4$ is thus given by
\begin{equations}
\label{eq: status t4}
H(t_4) &= H_0,\quad \rho_S (t_4)=\rho_S (t_3),\\
\rho_{A(B)}(t_4)&=\Tr_{B(A)} \prtg{\rot_S}=\rotr_{A(B)},\\
F_4 &= F(\rot_S,H_0).
\end{equations}
We remark that at $t=t_4$ the reduced states of  $A$ and $B$ (we name them reduced thermal states) are different from the initial ones, which were the local thermal states.

The turning on and off of the interaction requires work from system $R$ while, in general, $R$ is not involved during the thermalization from $t_2$ to $t_3$.
In particular, turning on the interaction costs a quantity $W_R (t_1 \rightarrow t_2)$, satisfying (more details in Appendix~\ref{sec: Work expense of system R}):
\begin{equation}\label{eq:work expense R on}
W_R (t_1 \rightarrow t_2) \geq F_2 - F_3 = \ev{H_I}_{t_2} + F_1 - F_3.
\end{equation}
On the other hand, turning it off costs:
\begin{equation}\label{eq:work expense R off}
W_R (t_3 \rightarrow t_4) \geq F_4 - F_1 = -\ev{H_I}_{t_3} + F_3 - F_1.
\end{equation}
Then, the minimum amount of work required to system $R$ to make one cycle is $\ev{H_I}_{t_2} -\ev{H_I}_{t_3} = -\Delta \ev{H_I}$. We stress that, during the switchings, system $R$ could lose a certain quantity of free energy due to dissipative effects, in addition to the required variations of usefulness.
In the following, we identify the amount of usefulness lost by $R$ with the variation of its free energy because we assume that the Hamiltonian of system $R$ never changes [see Eq.~\eqref{eq: work definition no Hamiltonian change}].

From the point of view of single-shot work extraction, the protocol ends at time $t_4$.
In the next section, we quantify the amount of work extracted and the efficiency of this process.
In order to iterate the process using the same systems $A$ and $B$, one has to transfer the extracted work at time $t=t_4$ to an external storage system $C$.
We provide an example of how to do this at the end of Sec.~\ref{sec: Rabi model} for a specific model.
After the transfer ($t=t_5=t_4 + \delta_T$), $A$ and $B$ are still in contact with the bath and, after a while, they will be again in their thermal state.
Then, the protocol can be done again from the start.

\subsection{\label{subsec:Work and efficiency}Work and efficiency}

By definition, the extracted work $W$ is equal to $F_4 - F_1$:
\begin{equation}
\label{eq: work by definition}
W = \ev{H_0}_{t_4} - \ev{H_0}_{t_1} - k_B T \prtq{S(\rot_S)-S\prt{\rot_A\otimes \rot_B}}.
\end{equation}
$W$ is thus composed by two parts: one purely energetic and one of entropic nature.
The entropic term appearing in Eq.~\eqref{eq: work by definition} comes directly from the adopted definition of work. Its presence assures the validity of the second law of thermodynamics.
Especially for finite systems, the entropic part can become much more important than the energetic one for non-vanishing temperatures.
An example of this behavior is shown in Sec.~\ref{sec: Two interacting qubits}. At $T=0$, instead, $W$ is a simple difference of average energies.
We stress that the work done on a system quantifies the change of usefulness of the system, $\Delta F(\rho,H)$.
Then, $\Delta F(\rho,H)$ being a function of the state and of the Hamiltonian of the system, it does not depend on the actual evolution that took place.

The extracted work can be rewritten as
\begin{equation}
\label{eq: extracted work}
W= \Delta F\prt{\rotr_A,H_A}+ \Delta F \prt{\rotr_B,H_B}+k_B T S(A:B),
\end{equation}
where $S(A:B)=S\prt{\rotr_A}+S\prt{\rotr_{B}}-S\prt{\rho_S^{\textup{th}}}$ is the mutual information between $A$ and $B$ for the state $\rho_S^{\textup{th}}$, a real non-negative quantity \cite{Petruccione_book}.
In particular, the mutual information term quantifies the amount of correlations between the two subsystems and its behavior is strongly model-dependent.
In Eq.~\eqref{eq: extracted work}, the only non-local entropic term is the mutual information as opposed to the local terms $S\prt{\rotr_A}$ and $S\prt{\rotr_B}$.
Then, we can also define the local work
\begin{equation}
W_l = \Delta F\prt{\rotr_A,H_A}+ \Delta F \prt{\rotr_B,H_B},
\end{equation}
which in some cases could be the only accessible work after the protocol.
The inequality $W_l \leq  W$ holds, an already known result of information thermodynamics \cite{Maruyama2009}, meaning that the amount of extracted work benefits from the presence of correlations in the final thermal state.
As can be seen in Secs.~\ref{sec: Two interacting qubits} and~\ref{sec: Rabi model}, the difference between local and global work can be significant.

The quantity $\Delta F\prt{\rho_{A(B)}^{\textup{rth}},H_{A(B)}}$ can be written as~\cite{Lostaglio_Tesi}
\begin{equation}
    \Delta F\prt{\rho_{A(B)}^{\textup{rth}},H_{A(B)}} =k_B T S\prt{\rho_{A(B)}^{\textup{rth}}||\rho_{A(B)}^{\textup{th}}},
\end{equation}
 where $S(\rho||\sigma) = \Tr \prtg{\rho \ln \rho} - \Tr \prtg{\rho \ln \sigma}$ is the relative entropy which, even not having all the properties of a distance measure, is often used to quantify how much two density operators are different \cite{Petruccione_book}.
Therefore, the more the reduced thermal states are different from the local ones, the more the local extracted work should be.
One then expects that $W_l$ should typically increase as the strength of the interaction between the subsystems of $S$ increases.

Another useful way to express the extracted work is through the partition functions of the systems.
Calling $Z_{A(B)}$ the partition function of system $A(B)$ with Hamiltonian $H_{A(B)}$ and $Z_S$ the partition function of the total system with interaction on, we can write:
\begin{equation}\label{eq:workandZ}
W=k_B T \ln \prt{\frac{Z_A Z_B}{Z_S}} - \ev{H_I}_{t_3}.
\end{equation}

Through simple algebraic manipulations, we can write
\begin{equation}
W=F_3-F_2-\Delta \ev{H_I},
\end{equation}
where we recall that $\Delta \ev{H_I}= \ev{H_I}_{t_3} - \ev{H_I}_{t_2}$.
Then,
\begin{equation}\label{eq:local-nonlocal-disequality}
0 \leq W_l \leq W \leq -\Delta \ev{H_I},
\end{equation}
because $F_3$ has to be always lower or equal to $F_2$.

Using Eq.~\eqref{eq:workandZ}, it is easy to show that if both subsystems are finite, the high-temperature limit of the extracted work is zero (see Appendix~\ref{sec: High temperature limit of extracted work in finite systems}).
This also implies, using Eq.~\eqref{eq:local-nonlocal-disequality},  that the correlations between two finite subsystems in a thermal state always go to zero faster than $1/T$ in the high-temperature limit since $W \rightarrow 0$.

Following the theorems of TRT, it is in principle always possible to transfer, without losses, a certain quantity of free energy from one system to another one through thermal operations.
Achieving the maximum efficiency for this transfer may require, for example, the use of catalysts \cite{WorkDef,Lostaglio_Tesi,Horodecki2013}.
Thus, we define the ideal efficiency of the process as the work stored in system $S$ divided by the minimum free energy lost by system $R$, i.e.,
\begin{equation}\label{eq:efficiency}
\eta = \frac{W}{-\Delta \ev{H_I}} = \frac{F_3-F_2-\Delta \ev{H_I}}{-\Delta \ev{H_I}}  \leq 1.
\end{equation}
In other words, we compare the work that system $S$ gains with the work that system $R$ would lose in the best-case scenario.
This comparison makes sense because TRT assures us that there exists a thermal operation such that all the work lost by $R$ is gained by $S$.
Of course, considering the local work $W_l$, with the annexed efficiency $\eta_l$, instead of $W$, we get $\eta_l \leq \eta$.

As can be seen from Eq.~\eqref{eq: entropy production}, the ideal efficiency of this process can be thought of depending explicitly on the entropy production of the thermalization, from time $t_2$ to $t_3$. With respect to the ideal switching case (Eq.~\eqref{eq:efficiency}), system $R$ could spend more work during the process because of dissipative effects.

It is worth noting that the extraction of work from $R$ to $S$ may imply the conversion of different forms of energy. Indeed, $R$ could exploit any kind of possible form of energy to switch on
and off the interaction between $A$ and $B$, while the form of energy stored in $S$ would depend on the specific choice of subsystems $A$ and $B$.

As an example of protocol implementation, we could think of a flying  atom entering and exiting from a cavity. In this case, the internal levels of the flying  atom are system $A$, the cavity is system $B$ and
the wave function spatial part of the  flying  atom is the resource $R$. When the atom enters or exits, the $A$-$B$ interaction  switches on or off and energy can come from or go to $R$. The amount of work paid by system $R$ may also depend on entropy variations and overall must be positive (see Eqs.~\eqref{eq:work expense R on} and~\eqref{eq:work expense R off}). In this specific example, mechanical energy is transformed into electromagnetic energy.

In order to make our analysis more quantitative and better exemplify the level of efficiency of our work-extraction protocol, we consider two possible realizations associated to two different models which can be realized in specific physical scenarios of experimental interest. In the next two sections we consider a two-qubit system and a spin-boson system described by the Rabi model.


\section{\label{sec: Two interacting qubits}Two interacting qubits}

Here, we consider the case when $S$ consists of two qubits governed by the Hamiltonian~\cite{Dawson2004}
\begin{equation}
H=H_A+H_B+H_I,
\end{equation}
where
\begin{equation}
H_A=\frac{\hbar \omega}{2}\sigma_z^{(A)},\
H_B=\frac{\hbar \omega}{2}\sigma_z^{(B)},\
H_I=\hbar g \sigma_x^{(A)}\sigma_x^{(B)},
\end{equation}
$\omega$ is the frequency of each qubit, $\sigma_z^{A(B)}$ and $\sigma_x^{A(B)}$ are Pauli matrices and $g$ is the coupling frequency.

The extracted work and the efficiency for this model can be computed by using Eqs.~\eqref{eq:workandZ} and \eqref{eq:efficiency} where the partition functions  and the average interaction energy at time $t_3$ (obtained through lengthy but straightforward calculations)  are given by
\begin{equations}
Z_{A}=Z_B&=2\cosh (\beta \hbar \omega/2),\\
Z_S &= 2 \prtq{\cosh \prt{\beta \hbar \sqrt{\omega^2 + g^2}}+\cosh \prt{\beta \hbar g}},
\end{equations}
and
\begin{eqnarray}
\nonumber \ev{H_I}_{t_3}
=&&\,\frac{\hbar g}{Z_S}\left[- \frac{2 e^{\beta\hbar \sqrt{\omega^2 + g^2}}}{g \N_-}\prt{\sqrt{\omega^2 + g^2} + \omega} -e^{\beta\hbar g} \right. \\ &&\, \left.
+e^{-\beta\hbar g}+\frac{2 e^{-\beta\hbar \sqrt{\omega^2 + g^2}}}{g \N_+}\prt{\sqrt{\omega^2 + g^2} - \omega}\right],\qquad
\end{eqnarray}
where $\N_{\pm} = \frac{2}{g^2}\prt{\omega^2 + g^2 \mp \omega \sqrt{\omega^2 + g^2}}$, while $\ev{H_I}_{t_2}=0$.

\begin{figure}[t!]
	\centering
	\includegraphics[width=0.48\textwidth]{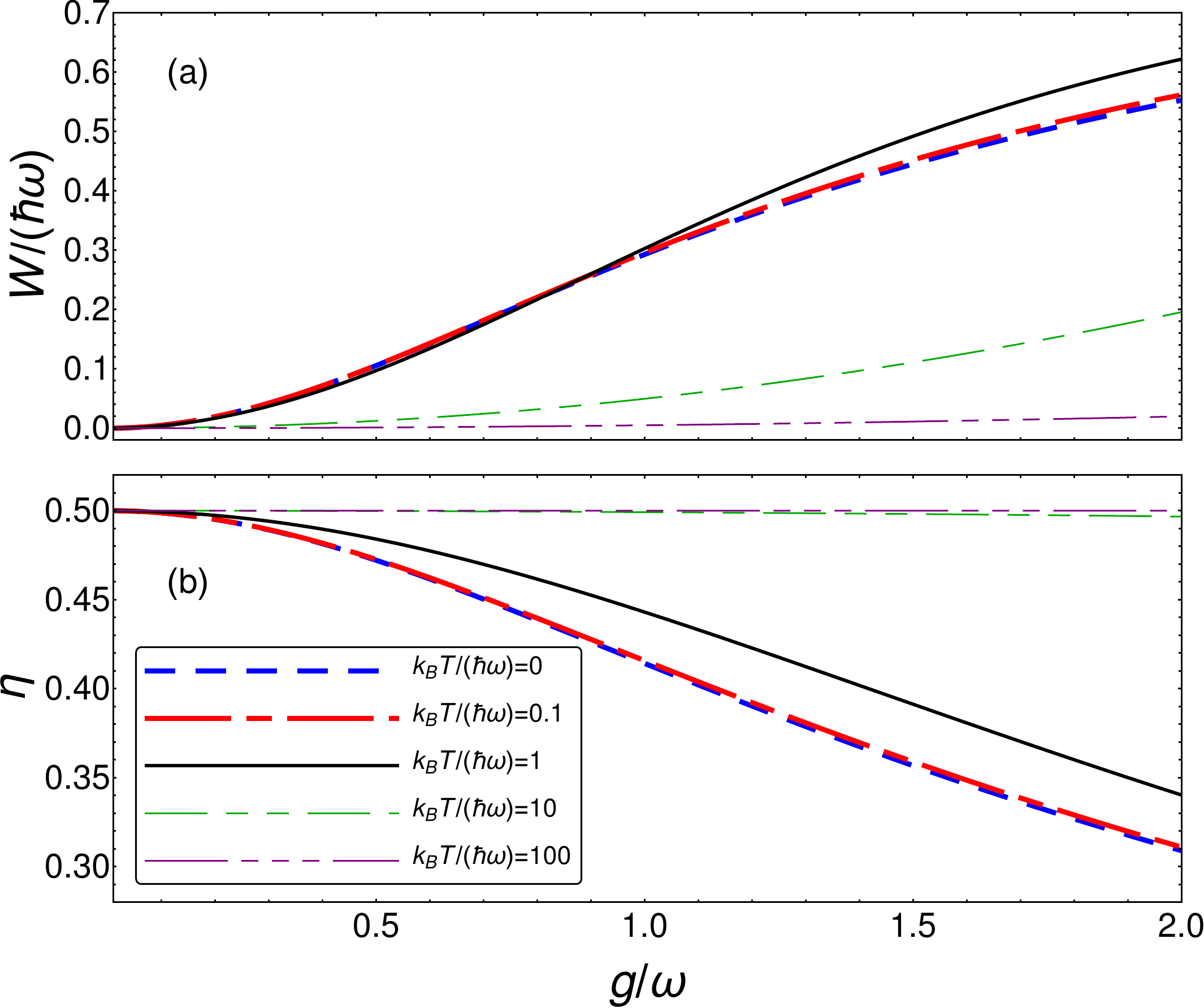}
	\caption{Two-qubit model: extracted work $W$ (part $(a)$) and efficiency $\eta$ (part $(b)$) of the thermalization protocol as a function of the coupling constant, $g/\omega$, for different values of the temperature of the bath, $k_B T/(\hbar\omega)$.
	}
	\label{fig: graphLavEffGlobInt}
\end{figure}

At zero temperature,  the extracted work and the efficiency assume the simple form
\begin{equations}
\label{eq: Zero temperature two qubits}
W(T=0)
&= \hbar\omega \frac{g^2}{\omega^2 + \omega\sqrt{\omega^2 + g^2}+g^2},\\
\eta (T=0)
&= \frac{\omega}{\omega + \sqrt{\omega^2 + g^2}}.
\end{equations}

Another analytical limit worth mentioning is the $g\rightarrow \infty$ limit.
In this case we obtain
\begin{equation}
W\rightarrow k_B T \ln \left\{1 + \cosh \prtq{\hbar\omega/(k_B T)}\right\},
\quad
\eta \rightarrow 0.
\end{equation}

\begin{figure}[t!]
	\centering
	\includegraphics[width=0.48\textwidth]{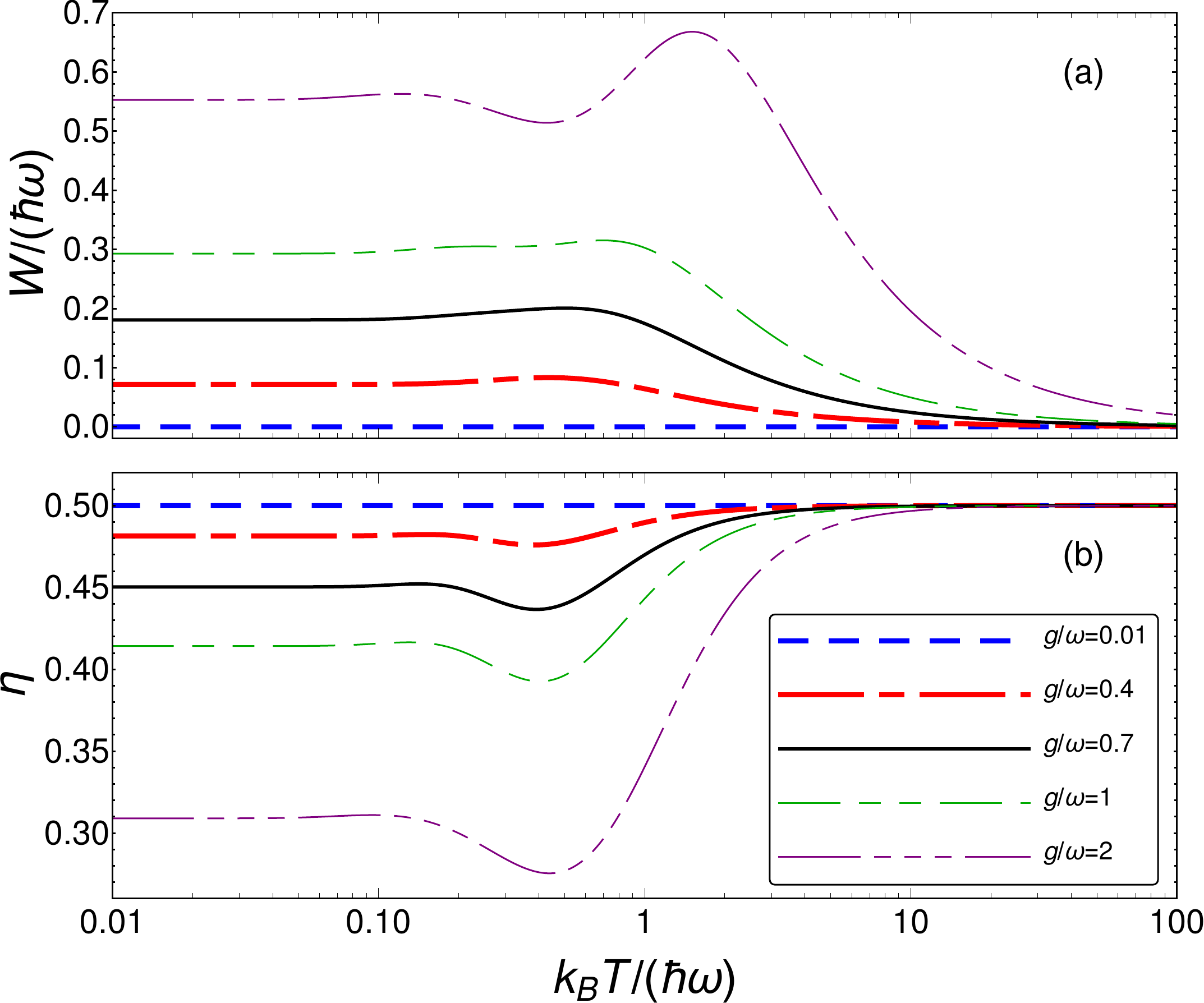}
	\caption{Two-qubit model: extracted work $W$ (part $(a)$), and  efficiency $\eta$ (part $(b)$) of the thermalization protocol as a function of the temperature of the bath, $k_B T/(\hbar\omega)$, for different values of the coupling constant, $g/\omega$.
	}
	\label{fig: graphLavEffGlobTemp}
\end{figure}

The behaviors of $W$ and $\eta$  as a function of the dimensionless coupling constant $g/\omega$ are plotted in Fig.~\ref{fig: graphLavEffGlobInt}, for  different temperatures of the bath.
We notice that for every temperature increasing $g/\omega$ monotonically increases the extracted work.
This behavior agrees with what was predicted in Sec.~\ref{subsec:Work and efficiency} for the $T=0$ case ($W=W_l$). An extraction of work comparable with the typical energies of the subsystems can be obtained.
However, the efficiency also decreases monotonically.
Then, for a given $T$, a sweet spot for the coupling constant does not seem to exist.
In contrast, a sweet spot for the temperature does exist.
Indeed, as shown in Fig.~\ref{fig: graphLavEffGlobTemp}, the most interesting feature of this model is that, given a value of $g/\omega$, the maximal extraction of work is obtained for a value of temperature such that $k_B T \sim \hbar\omega$ with a greater efficiency with respect to the zero-temperature case.

\begin{figure}[!t]
	\centering
	\includegraphics[width=0.48\textwidth]{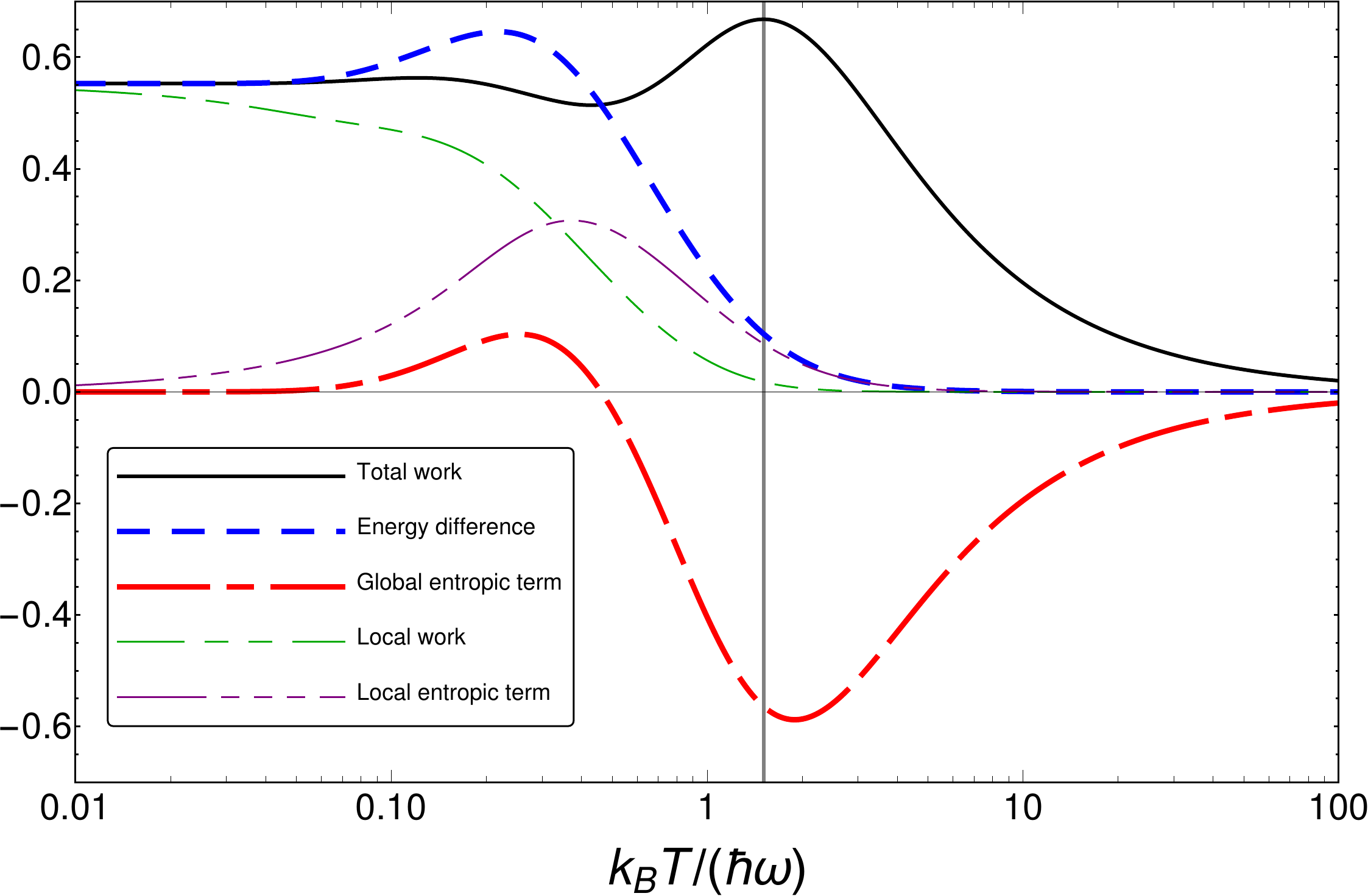}
	\caption{Two-qubit model: comparison of different quantities (each in units of $\hbar \omega$) as a function of the bath temperature for  $g=2  \ \omega$. A gray vertical line is depicted in correspondence of the peak of the total work $W $at $k_B T \simeq 1.51  \ \hbar\omega$.}
	\label{fig: graphConfrontoG2}
\end{figure}

The above result is especially clear for high values of $g$ such as $g=2 \ \omega$.
For this reason, we plot in Fig.~\ref{fig: graphConfrontoG2} various quantities of the protocol as a function of the temperature, for $g=2 \ \omega$.
The maximal extraction of total work $W$ is obtained for $k_B T \simeq 1.51  \ \hbar\omega$, marked with a gray vertical line in the figure.
Around that temperature, there is a big difference between the total work $W$ and the local one $W_l$.
Their difference is exactly the mutual information multiplied by $k_B T$.
So, a great part of the work is stored in the non-local entropic term $ k_B T S(A:B)$.
Regarding the total extraction, we can notice also how much the global entropic term, $k_B T \prtq{S\prt{\rot_S}-S\prt{\rot_A\otimes \rot_B}}$, is important in that temperature region.
In contrast, for lower values of $T$, the global entropic term reduces the amount of extracted work with respect to the energy difference $\ev{H_0}_{t_4} - \ev{H_0}_{t_1}$.
On the local level, the local entropic term, $k_B T \prtq{S\prt{\rho_{A}^{\textup{rth}}\otimes \rho_{B}^{\textup{rth}}}-S\prt{\rot_A\otimes \rot_B}}$, always reduces the amount of work extracted, independently of the temperature.
This difference of behavior between the local and non-local parts of the entropy explains the quantitative difference between the local and total extracted work.

\section{\label{sec: Rabi model}Rabi model}

\begin{figure}[t!]
	\centering
	\includegraphics[width=0.48\textwidth]{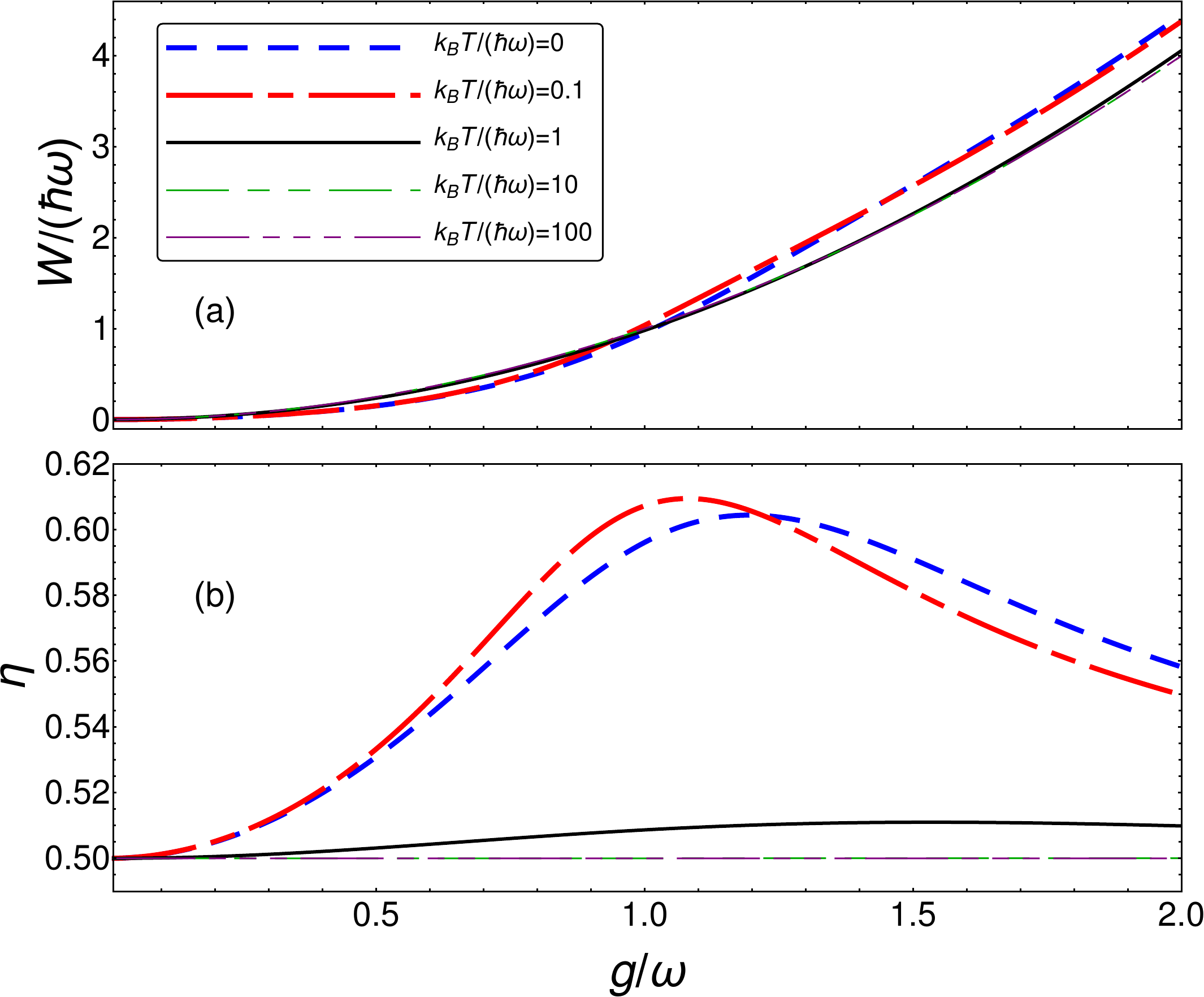}
	\caption{Rabi model: extracted work $W$ (part $(a)$) and efficiency $\eta$ (part $(b)$) of the thermalization protocol as a function of the coupling constant, $g/\omega$, for different values of the bath temperature, $k_B T/(\hbar\omega)$.	}
	\label{fig: graphRabiLavEffGlobInt}
\end{figure}

Here, we consider the case in which $S$ consists of a two-level system (subsystem $A$) interacting with a harmonic oscillator (subsystem $B$).
The system is governed by the Rabi Hamiltonian \cite{Haroche_book}:
\begin{equation}\label{eq: RabiHamiltonian}
H_{\textup{Rb}} =H_A+H_B+H_I,
\end{equation}
where
\begin{equation}\label{eq: RabiHamiltonian2}
H_A =\hbar\Delta \sigma_z,\quad
H_B=\hbar \omega \hat{n},\quad
H_I=\hbar g \sigma_x \prt{a^\dagger + a},
\end{equation}
$\hbar\Delta$ is half of the energy distance between the ground state $\ket{g}$ and the excited state $\ket{e}$ of $A$, $\omega$ is the frequency of $B$ (typically $\omega \sim 2\Delta$), $\hat{n}$ is the number operator (with the number basis given by $\hat{n}\ket{n}=n\ket{n}$), $a^\dagger$ and $a$ are the creation and annihilation operators, and $\sigma_z$ and $\sigma_x$ are the Pauli matrices.

Since the analytical solution of the Rabi model is given in terms of series that have to be truncated \cite{Braak,Quantum_Rabi_Review,QRM_Bogoliubov},
from a numerical point of view, it is easier to directly do all the computations numerically without using the analytical solution.
In this section, we report the results of these numerical simulations done with the PYTHON package QUTIP \cite{Johansson2012,Johansson2013}, only dealing with the resonant case $\Delta=\omega/2$.
For the zero-temperature case, we use the analytical solution, checking that it coincides  with the numerical simulations at very low temperatures.
A  detailed discussion about the ground state of the system, used for the $T=0$ case, can be found in Appendix~\ref{sec: Study of the Rabi Hamiltonian ground state}.

Fig.~\ref{fig: graphRabiLavEffGlobInt} shows the extracted work and the efficiency as a function of the coupling parameter $g/\omega$ for different values of the bath temperature.
We notice a dissimilar behavior of the Rabi model with respect to the two-qubit one.
Overall, the Rabi model attains a higher value of extracted work and higher efficiency.
In contrast to the two-qubit case, here increasing the interaction may increase the efficiency, which is always higher than $1/2$.

\begin{figure}[!t]
	\centering
	\includegraphics[width=0.48\textwidth]{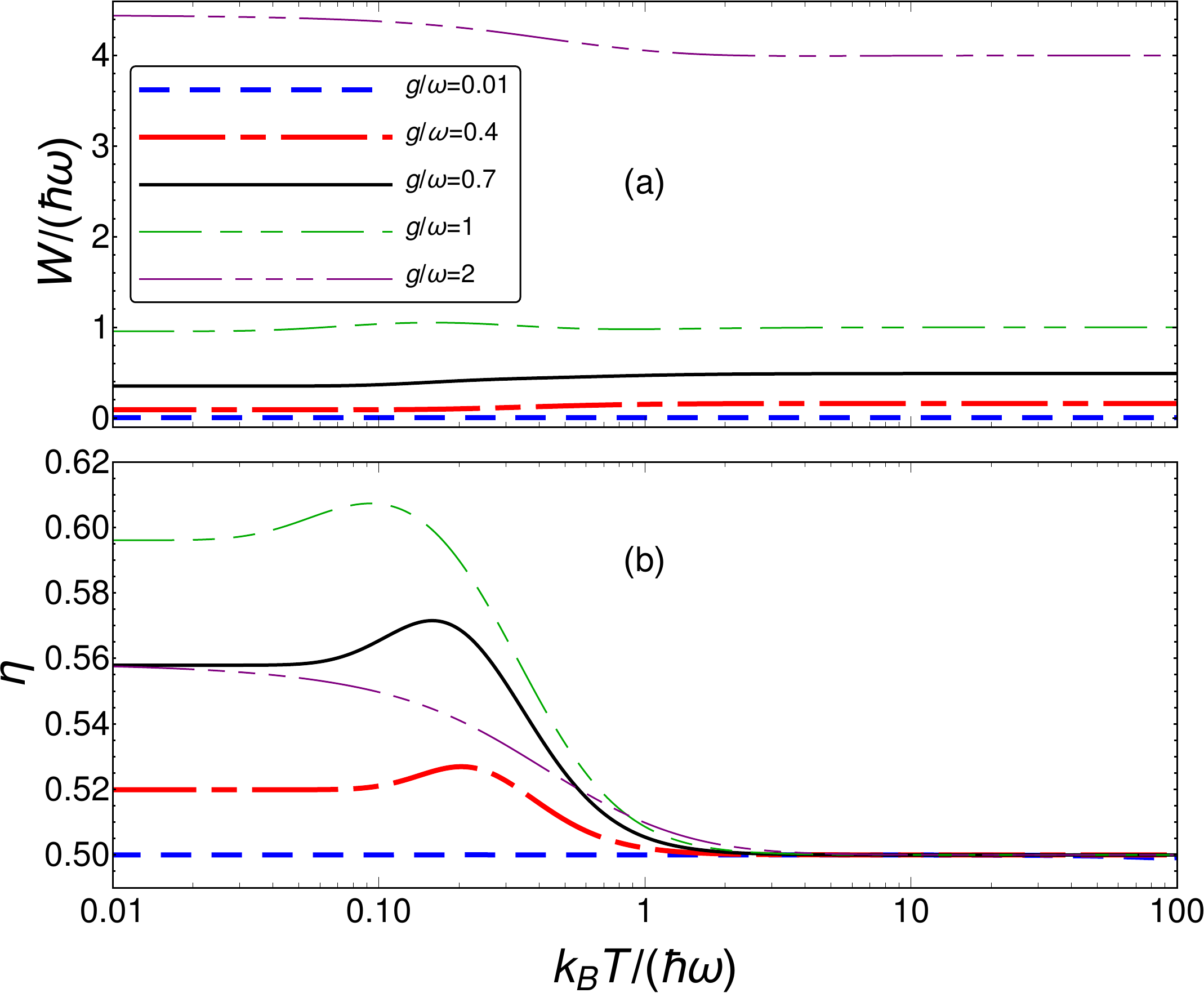}
	\caption{Rabi model: extracted work $W$ (part $(a)$) and efficiency $\eta$ (part $(b)$) of the thermalization protocol as a function of the bath temperature, $k_B T/ (\hbar\omega)$, for different values of the coupling constant, $g/\omega$.}
	\label{fig: graphRabiLavEffGlobTemp}
\end{figure}

As in the two-qubit case, also in the Rabi model, for a given value of the coupling constant, an ideal value of the temperature exists (see Fig.~\ref{fig: graphRabiLavEffGlobTemp}).
Comparing Figs.~\ref{fig: graphLavEffGlobTemp} and \ref{fig: graphRabiLavEffGlobTemp} we notice that the best values of temperature in the Rabi model case are one order of magnitude lower. In general, given a value of the coupling constant $g$, there exists a temperature sweet spot where the work is nearly at its maximum and, close to which, the efficiency has its peak.
Among the values reported in the plot, this does not hold for $g= 2 \ \omega$.
Moreover, the extracted work does not tend to zero as in the two-qubit case.
This is due to the fact that the Rabi Hamiltonian contains a non-finite and nonbounded system (the harmonic oscillator).
This means that a  temperature that makes all the levels equally populated so that the thermal state is practically the identity state does not exist. We also remark that both the work and the efficiency reach an asymptotic behavior for $k_B T \sim 10 \ \hbar\omega$.

\begin{figure}[!t]
	\centering
	\includegraphics[width=0.48\textwidth]{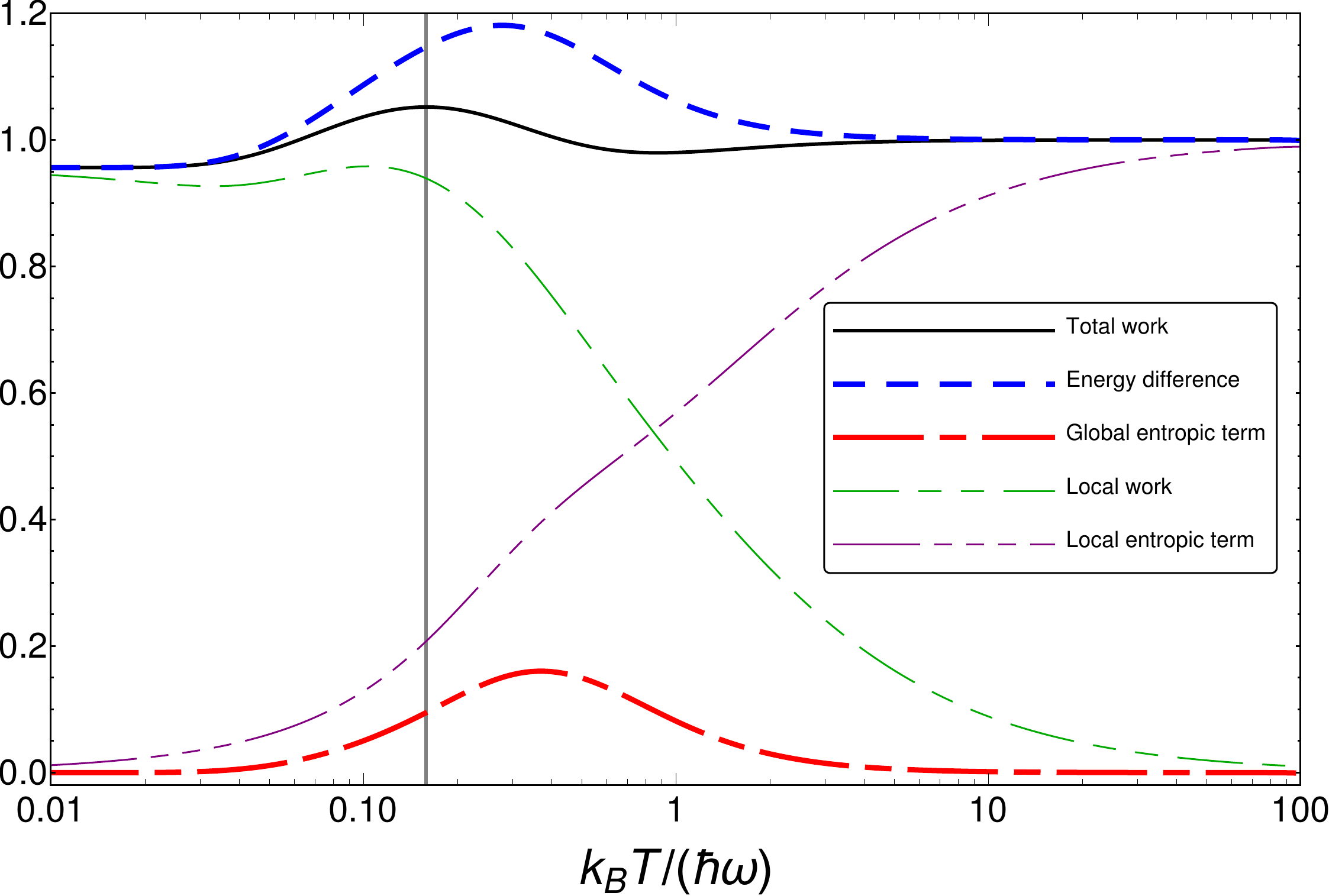}
	\caption{Rabi model: comparison of different quantities (each in units of $\hbar \omega$) as a function of the bath temperature for  $g=\omega$.
	A gray vertical line is depicted in correspondence of the peak of the total work $W $at $k_B T \simeq 0.16\ \hbar\omega$.
	}
	\label{fig: graphRabiConfronto}
\end{figure}

In Fig.~\ref{fig: graphRabiConfronto} we show the extracted work and other relevant quantities as a function of the temperature (cf. Fig.~\ref{fig: graphConfrontoG2}).
Differently from the two-qubit case, here the peak of work extraction is not due to the entropic term but to the energy term.
Indeed, the peak of work extraction (roughly at $k_B T \simeq 0.16\ \hbar\omega$ and marked with a gray vertical line in the figure) is near the peak of the energy difference term.
As in the two-qubit case, the local work rapidly goes to zero starting from $k_B T\sim \hbar\omega$.
In the present case, however, this is not due to the fact that the reduced thermal states are very similar to the thermal ones.
In this case, the energy difference remains high and the local entropic term counterbalances it.
Then, even if the global entropic term does not seem to play a significant role, its non-local component (the mutual information) does, by balancing the local entropic terms and thus avoiding that they take the total work down to zero in the high-temperature region.

{\bf Charging a battery. --- } In general, the aim of work-extraction protocols is to realize something useful, e.g. a process that could be seen as the \emph{charge of a battery}. Here, we propose a \emph{transfer protocol} (for the zero-temperature case) that allows for storing the energy gained by system $S$ into an external harmonic oscillator (system $C$) which plays the role of a battery, the energy of which can be increased more and more through suitable cyclic interactions with system $S$.

The main idea is to imagine that system $A$ is not just a two-level system, but a three-level system the intermediate level of which does not participate in the interaction with system $B$.
However, system $C$ is resonant with the transition from this intermediate level to the excited one and interacts with system $A$ through a Jaynes-Cummings Hamiltonian.
In this way, whatever is the population of the ground level of system $A$, it can only give energy to system $C$ but not receive it.
Moreover, we use, as free resources, a certain number of systems having the same spectrum of system $A$, to transfer energy from $B$ to $C$.
Imagining the role of system $R$ being played by the wave function spatial part of each three-level system approaching and then leaving the harmonic oscillator (one can think of flying atoms and a cavity), we can clearly see that we cannot charge system $C$ by using directly the three-level systems in their initial state at $T=0$ (ground state), i.e.,  without first charging them through their interaction with system $B$. We observe that in this specific case, then, the resource would have a genuine quantum nature.
An extensive description of this transfer protocol is reported in Appendix~\ref{sec: An example of transfer protocol}.

In a specific simulation of this transfer protocol (in the deep-strong coupling regime, $g > \omega$, and for $100$ iterations) we obtain a final energy transferred to system $C$  of the order of $100 \hbar \omega$ with a low standard deviation and a  reasonable efficiency (see table \ref{table} of Appendix~\ref{sec: An example of transfer protocol}).

\section{\label{sec:Conclusions}Conclusions}

In this paper, we have proposed a protocol of work extraction from a resource system $R$ to a storage bipartite system $S$, based on the thermalization of the latter.
We first described the protocol in the general case without referring to a specific implementation.
This thermalization protocol should be easily implementable because the only requirement is to turn on and off the interaction between subsystems of $S$ ($A$ and $B$) in a short amount of time without changing the state of system $S$.
Results from thermodynamic resource theory have been then used to define the extracted work $W$ and the efficiency $\eta$.

Then, we applied this protocol to two models:
a simple two-qubit system and a system described by the Rabi Hamiltonian.
In both cases, we pointed out the important role that the entropic terms play when the protocol is applied at temperatures comparable with the energies of the subsystems and the great difference between local and total extracted work.
In both cases, the extracted work is comparable with the frequencies of the systems.
Moreover, we remark that in the Rabi model case the efficiency we have obtained is always higher than one half.

We have thus pointed out that simple processes like thermalization and the switching on and off of interactions between quantum systems can be exploited as a potential resource for thermal machines.

Finally, as a proof of principle, we showed how the energy extracted after one cycle of the thermalization protocol (single-shot extraction protocol) can be transferred to an external work storage system through a transfer process which makes the complete protocol iterable.
We stress that the whole protocol, composed by many iterations of the thermalization protocol plus the transfer protocol, realizes something like the \emph{charge of a battery} in a realistic scenario, not involving fine operations.

We believe it would be interesting to generalize our study of the Rabi model to the case of $N$ qubits interacting with a harmonic oscillator.
This would allow one to study if the correlations that would be built among the qubits allow for a greater work extraction.
Finally, further studies could suggest how to improve our transfer protocol and pave the way to proposals in specific physical scenarios.

\section*{Acknowledgements}

N.P. acknowledges the financial support of the Erasmus+ program of the European Union and of the Institut UTINAM for the development of this program. N.P. and B.B. acknowledge useful suggestions by Felipe Fernandes Fanchini and thank the IT team of the Institut UTINAM  for its
technical support.

\appendix

\section{\label{sec: Principles of thermodynamic resource theory} Thermodynamic resource theory}

In general, a resource theory is a theory in which the possible operations that can be done on a system are restricted by some constraints and it is supposed that to perform a given protocol some special states belonging to other systems are unlimitedly available for free.
These external systems in these particular states are called \enquote{free resources} \cite{Lostaglio_Tesi}.
An operation $\mathcal{T}(\rho)$ permitted by TRT and addressed as a thermal operation has the following structure:
\begin{equation}
 \mathcal{T}(\rho)=\Tr_{\textup{bath}} \prtg{U (\rho \otimes \rho_{\textup{bath}})U^\dagger},  \quad\comm{H+H_{\textup{bath}}}{U}=0,
\end{equation}
where $H$ is the Hamiltonian of the system described by the density operator $\rho$, $H_{\textup{bath}}$ is the Hamiltonian of the free resource (usually called the bath), and $U$ is a generic unitary operator that connects initial and final states of the total system (free resource included).
The free resource (bath) is a system with an arbitrary Hamiltonian, assumed to be in a thermal state ($\rho_{\textup{bath}}$) at a given temperature.
In a thermal operation the bath is used only for the duration of the protocol
so that there is no interaction between system and bath at the start and at the end of the protocol.
Then, the commutator $\comm{H+H_{\textup{bath}}}{U}=0$ assures us that the global system has kept its energy unchanged (not only the mean value but also the entire distribution).
This is needed because the aim of TRT is to keep track of all the energy terms involved in a possible thermal process and to find what bounds this constraint generates.

In Appendix~\ref{sec: Work expense of system R} we also consider an extended class of permitted operations, called catalytic Gibbs-preserving transitions~(see Ref. \cite{WorkDef} for their definition), which are used to analyze the switching on and off of the interaction Hamiltonian between subsystems $A$ and $B$.
The important property that we use is that the quantity $\Delta F(\rho,H)$  cannot increase under such transitions.

\section{\label{sec: Work definition, heat and entropy production}Work definition, heat and entropy production}

Here, we make some comments on what could be the consequences of the adopted work quantifier on the definition of heat and on its connection with the entropy production. Let us assume that we can use $\Delta U = \Tr\prtg{H'\rho'}-\Tr\prtg{H \rho}$ as the internal energy change in the first law of thermodynamics, $\Delta U = Q + W$. It follows that
\begin{multline}\label{eq:Q}
	Q= \Tr\prtg{H' {\rho'}^{\textup{th}}} - \Tr\prtg{H \rot} +\\
	+ k_B T\prtq{S\prt{\rot }-S\prt{{\rho'}^{\textup{th}}} + S\prt{\rho'}-S(\rho)},
\end{multline}
where  $\rho^{\textup{th}}$ and ${\rho'}^{\textup{th}}$ are the thermal states corresponding, respectively, to $H$ and $H'$.
When only the system state changes we have:
\begin{equation}
	\Delta U = \Tr\prtg{H \prt{\rho'-\rho}}, \quad
	Q = k_B T \prtq{S\prt{\rho'}-S(\rho)},
\end{equation}
which seems reasonable as the heat is given by  the change of entropy times $ k_B T$, $T$ being the temperature at which the process takes place.
On the other hand, when only the Hamiltonian changes while the state of the system does not, $\Delta U$ and $ Q$ reduce to:
\begin{equations}
	\Delta U &= \Tr\prtg{\prt{H'-H}\rho}, \\
	Q &=F\prt{{\rho'}^{\textup{th}},H'} - F\prt{\rot,H}.
\end{equations}

Let us comment now on a possible connection with the entropy production.
During a thermal operation, system and bath together evolve unitarily so that the total entropy of both systems does not change and we can apply findings of Ref. \cite{Esposito2010}.
There, the system under analysis is unitarily interacting with one or more thermal baths.
To adapt the equations to our case we will use one single thermal bath.
In particular, we focus on a generic time interval with time-independent Hamiltonians, as from $t_2$ to $t_3$ in our thermalization protocol.

The system and the bath are considered to be in the state $\rho (0) = \rho_S (0) \otimes \rho_\bath^\textup{th}$ at time $t=0$, as we also assume in our case during the thermalization step (with $t_2$ in place of $t=0$).
In particular, the entropy change in the system during the evolution can be decomposed as follows:
\begin{equation}
\Delta S (t) = \Delta_i S (t) + \Delta_e S(t),
\end{equation}
where $\Delta_i S (t)$ is the entropy production and $\Delta_e S(t)$ represents the reversible contribution to the system entropy due to heat exchanges.
More specifically~\cite{Esposito2010}, $\Delta_e S (t) = \beta Q_\bath (t)$, where $Q_\bath (t) \equiv \ev{H_\bath}_{t=0} - \ev{H_\bath}_{t}$ represents the heat flow from the reservoir [here $\beta=1/(k_B T)$].

We are only interested in what happens at the end of the thermalization protocol, where TRT imposes the conservation of the total energy (see Appendix~\ref{sec: Principles of thermodynamic resource theory}), therefore in this case $Q_\bath (t)$ is equal to $\Delta \ev{H_S}_t = \Tr\prtg{H_S \rho_S (t)} -\Tr\prtg{H_S \rho_S (0)}$.
We recall that free energy is a decreasing monotone of thermal operations, that is $F\prt{\rho_S(0),H_S}-F\prt{\rho_S (t),H_S} \geq 0 $.
Then, we can show that $\Delta_i S (t) \geq 0$ as follows:
\begin{equations}\label{eq: entropy production}
	\Delta_i S (t)
	&= -\beta Q_\bath(t) + \Delta S(t)\\
	&=\beta\prtq{-\Delta \ev{H_S}_t + \frac{1}{\beta} \Delta S (t)}\\
	&=\beta\prtq{F(\rho_S(0),H_S)-F(\rho_S(t),H_S)}\geq 0.
\end{equations}
Notice that $\Delta_i S (t) = - \beta W (t)$, i.e., the entropy production exactly matches the loss of \enquote{usefulness}  of  system $S$ times the inverse temperature of the environment.

In the other steps of the protocol no entropy is produced because we assume ideal switchings (i.e., reversible processes) to define the ideal efficiency. Of course,  some entropy is expected to be produced in a realistic implementation even during these operations.

\section{\label{sec: Work expense of system R}Work expense of system R}

Here, we show the amount of free energy that system $R$ has to lose to turn on the interaction of system $S$.
Considering the whole system $S+R$, before the action of system $R$ we have:
\begin{equation}
\Delta F \prt{\rho_R \otimes \rot_{AB}, H_R + H_0} = \Delta F \prt{\rho_R, H_R},
\end{equation}
where $\rot_{AB}=\rot_A\otimes \rot_B$.
After the action of $R$, we have:
\begin{multline}
\Delta F \prt{\rho'_{RAB},H_R+H_0+H_I} \geq\\
\geq \Delta F \prt{\rho'_R,H_R}+\Delta F \prt{\rot_{AB},H_0 + H_I}.
\end{multline}
We consider the operation under consideration to be a catalytic Gibbs-preserving transition so that $\Delta F \prt{\rho, H}$ has  to decrease or to stay constant~\cite{WorkDef}, therefore:
\begin{equation}
F \prt{\rho'_R, H_R} - F \prt{\rho_R, H_R} \leq \Delta F \prt{\rot_{AB} ,H_0+ H_I}.
\end{equation}
In the above equation, the equality holds in the best-case scenario.  The work expense of $R$ to perform the switching on is then given by Eq.~\eqref{eq:work expense R on} while,
analogously, one can obtain Eq.~\eqref{eq:work expense R off}  for the switching off.

\section{\label{sec: High temperature limit of extracted work in finite systems}High temperature limit of extracted work in finite systems}

If both  $A$ and $B$  are finite, in the high-temperature limit ($\beta \rightarrow 0$) at first order in $\beta$ the following expansion holds:
\begin{equation}
e^{-\beta H} \simeq I - \beta H,
\end{equation}
where $H$ is the Hamiltonian of the whole bipartite system and $I $ is the identity in the whole Hilbert space.
We call $N_A$ the dimension of subsystem $A$ and we call $N_B$ the dimension of subsystem $B$ while $N_S=N_A N_B$.
We use Eq.~\eqref{eq:workandZ} written in the following way:
\begin{equation}
W= \frac{1}{\beta}\ln \prt{Z_A Z_B}-\frac{1}{\beta}\ln(Z_S) - \ev{H_I}_{t_3}.
\end{equation}
Expanding up to first order in $\beta$ we get
\begin{equations}
Z_{A(B)} & \simeq N_{A(B)} - \beta \Tr_{A(B)}\prtg{H_{A(B)}},\\
Z_A Z_B &\simeq N_S - \beta\prt{N_A \Tr_B\prtg{H_B}+N_B\Tr_A\prtg{H_A}},\\
\ln(Z_A Z_B) &\simeq \ln N_S - \frac{\beta}{N_S}\left(N_A \Tr_B\prtg{H_B} \right.\\
&\quad  \left.+N_B\Tr_A\prtg{H_A}\right).
\end{equations}
Similarly
\begin{equations}
Z_S &\simeq [Z_A Z_B]^{(1)} - \beta \Tr_S\prtg{H_I},\\
\ln(Z_S) &\simeq [\ln\prt{Z_A Z_B}]^{(1)}  - \frac{\beta}{N_S}\Tr_S\prtg{H_I},
\end{equations}
where the terms $[Z_A Z_B]^{(1)} $ and $[\ln(Z_A Z_B)]^{(1)} $  are the functions between brackets computed at first order in $\beta$.
Lastly, to order zero in $\beta$:
\begin{equation}
\ev{H_I}_{t_3} = \Tr_S \prtg{H_I \frac{e^{-\beta (H_A+H_B+H_I)}}{Z_S}} \simeq \frac{\Tr_S\prtg{H_I}}{N_S}.
\end{equation}
Then, by considering all the contributions we obtain
\begin{equation}
\lim_{\beta \rightarrow 0} W = 0.
\end{equation}
This result also implies
\begin{equation}
\lim_{\beta \rightarrow 0} \frac{1}{\beta}S(A:B) = 0,
\end{equation}
because the global work is always higher than or equal to the local one, but they are both positive and their difference is given by the correlation term.

\section{\label{sec: Study of the Rabi Hamiltonian ground state}  Rabi Hamiltonian ground state}

In this Appendix, all the quantities with the tilde are in units of $\omega$ to lighten the notation ($\tilde{X}\equiv X/\omega$).

At $T=0$, the entropy terms do not contribute to the free energies and, then, we can deal with average energies only.
As a consequence, $W=W_l$, which, using $\ev{H_I}_{t_2}=0$ in Eq.~\eqref{eq: work by definition}, takes the form
\begin{equation}
\label{eq: work in QRM}
W=\ev{H_A}_{t_3}+\ev{H_B}_{t_3} + \hbar \Delta=\hbar \nu_0-\ev{H_I}_{t_3}+\hbar\Delta,
\end{equation}
where $\hbar \nu_0$ is the energy of the ground state of the Rabi model.
The efficiency of Eq.~\eqref{eq:efficiency} is given by
\begin{equation}
\label{eq: eff in QRM}
\eta=\frac{\ev{H_A}_{t_3}+\ev{H_B}_{t_3} + \hbar\Delta}{-\ev{H_I}_{t_3}}\,.
\end{equation}

In order to calculate the quantities in Eqs.~\eqref{eq: work in QRM} and \eqref{eq: eff in QRM} we need to study the ground state and how it is decomposed in the bare basis. To this end we  mainly follow the approach and the formalism of Ref.~\cite{Quantum_Rabi_Review}.
These calculations allow us to compute numerically, but starting from the formal and analytical solutions, the amount of extracted work and the efficiency of the protocol.

Following Ref.~\cite{Quantum_Rabi_Review}, the ground energy $\hbar \nu_0$ of the ground state of the Rabi Hamiltonian of Eqs.~\eqref{eq: RabiHamiltonian} and \eqref{eq: RabiHamiltonian2} can be calculated by searching for the first zero of Braak's function $G_-(x)$ \cite{Braak,Quantum_Rabi_Review,QRM_Bogoliubov}, defined by:
\begin{equation}
\label{eq: Braak function}
G_\pm (x) = \sum_{n=0}^{\infty} \left(1 \mp \frac{\tilde{\Delta}}{x-n} \right)  f_n \tilde{g}^n =0,
\end{equation}
where $x=\tilde{\nu}+\tilde{g}^2$.
The factors $f_n$ are calculated by recurrence through the following formulas:

\begin{equation}
\begin{aligned}
f_n &=\frac{1}{n}\prtq{\Omega (n-1)f_{n-1} - f_{n-2}}, \quad f_0=1, \ f_1=\Omega (0),\\
\Omega (n) &=\frac{1}{2 \tilde{g}}\prt{n + 3 \tilde{g}^2-\tilde{\nu}-\frac{ \tilde{\Delta}^2}{n-\tilde{g}^2-\tilde{\nu}}  }.
\end{aligned}
\end{equation}
The values of $\nu$ for which Braak's functions are zero are the eigenvalues of the Rabi Hamiltonian.
The lowest of these eigenvalues is the ground energy of the system.

According to \cite{Quantum_Rabi_Review}, after some easy but lengthy calculations the ground state can be written as follows:
\begin{equation}
\label{eq: the ground state}
\ket{\psi_g}=\frac{1}{2\sqrt{\mathcal{N}}}\prtq{\ket{e}(\ket{\phi_1}+\ket{\phi_2})+\ket{g}(\ket{\phi_1}-\ket{\phi_2})},
\end{equation}
where $\mathcal{N}$ is a normalisation constant,
\begin{widetext}
	\begin{equation}
	\label{eq: cavity reduction 1}
	\braket{n}{\phi_1} = e^{-\tilde{g}^2/2} \sqrt{n!} \sum_{m=0}^\infty m!\ e_m \left[ \sum_{k=\max(0,n-m)}^n \frac{(-1)^k}{[m-(n-k)]!(n-k)!k!} \tilde{g}^{m-(n-2k)} \right],\ e_m =-\frac{\tilde{\Delta}}{m- \tilde{g}^2-\tilde{\nu}_0}f_m,
	\end{equation}
\end{widetext}
and $\braket{n}{\phi_2}$ is equal to $\ip{n}{\phi_1}$ if one replaces in its expression $e_m$ with $f_m$.
We notice that both $\langle n|\phi_1 \rangle$ and $\langle n|\phi_2 \rangle$ are real.

The parity operator $\Pi~=~-\sigma_z (-1)^{\hat{n}}$ commutes with $H_{\textup{Rb}}$.
Thus, it is easy to show that the ground state of the Rabi Hamiltonian has to be of the form
\begin{equation}
\label{eq: form of ground state}
\ket{\psi_g}= \sum_{n=0}^{\infty} c_{2n} \ket{g,2n} + \sum_{n=0}^{\infty} c_{2n+1} \ket{e,2n+1}.
\end{equation}
Indeed, for low values of $\tilde{g}$ the ground state has to contain the component $\ket{g,0}$ so that all the other components have to be of the same parity.
Moreover, for every value of $\tilde{g}$ the ground eigenvalue does not cross with the others eigenvalues, therefore the ground state has the same parity for each value of $\tilde{g}$.
By taking the scalar product of both sides of \eqref{eq: the ground state} with $\ket{n}$, odd or even, and checking \eqref{eq: form of ground state} one can easily infer the following equalities:
\begin{equation}
\label{eq: parity of coefficients}
\sqrt{\mathcal{N}}c_n=\braket{n}{\phi_1}=(-1)^{n+1}\braket{n}{\phi_2}, \qquad \forall n.
\end{equation}

Now we can easily calculate the reduced states and write down the quantities of interest.
First of all, let us observe that
\begin{equation}\label{B state after therm}
\rho_B^{\textup{rth}}=\Tr_A \prtg{\dyad{\psi_g}}= \frac{1}{2\mathcal{N}}\left(\dyad{\phi_1} + \dyad{\phi_2} \right).
\end{equation}
Then, exploiting Eq.~(\ref{eq: parity of coefficients}), the average energy of the harmonic oscillator can be written as follows:
\begin{equation}
\ev{H_B}_{t_3}=\frac{\hbar\omega}{\mathcal{N}}\sum_{n=0}^\infty n\abs{\braket{n}{\phi_1}}^2.
\end{equation}

Similarly, the reduced state of the two-level system is found to be
\begin{widetext}
	\begin{equations} \label{A state after therm}
		\rho_A^{\textup{rth}}&=\Tr_B \prtg{\dyad{\psi_g}}=\frac{1}{\mathcal{N}}\left[\left(\sum_{n=0}^\infty \abs{\braket{2n+1}{\phi_1}}^2 \right)\dyad{e}+
		\left(\sum_{n=0}^\infty \abs{\braket{2n}{\phi_1}}^2 \right)\dyad{g}\right]\\
		&=\left(\sum_{n=0}^\infty \abs{c_{2n+1}}^2 \right)\dyad{e}+
		\left(\sum_{n=0}^\infty \abs{c_{2n}}^2 \right)\dyad{g},
	\end{equations}
\end{widetext}
and the average energy is
\begin{equations}
	\ev{H_A}_{t_3}=\hbar\Delta\prt{2\sum_{n=0}^\infty \abs{c_{2n+1}}^2-1}.
\end{equations}

Concerning the average of the interaction energy $\ev{H_I}_{t_3}$, it can be directly calculated with the formula:
\begin{equation}
\ev{H_I}_{t_3}=\frac{2\hbar g}{\mathcal{N}}\sum_{n=0}^\infty \sqrt{n+1}\braket{n+1}{\phi_1}\braket{n}{\phi_1},
\end{equation}
or, alternatively, it can be inferred from Eq.~(\ref{eq: work in QRM}) as we already know $\ev{H_A}_{t_3}$ and $\ev{H_B}_{t_3}$ which we calculated through the knowledge of $\nu_0$.

For very low values of $g/\omega$ the Jaynes-Cummings approximation can be used \cite{Petruccione_book}.
In this case Eq.~\eqref{eq: the ground state} becomes $\ket{\psi_g} \simeq \ket{g,0}$.
The Bloch-Siegert approximation holds well for higher values of $g/\omega$ (still $g/\omega \ll 1$) \cite{Bloch-Siegert}.
In this case, at resonance, $2\Delta=\omega$, Eq.~\eqref{eq: the ground state} becomes
\begin{equation}
\label{eq: ground state BS}
\ket{\psi_g}\simeq \left(1-\frac{\Lambda^2}{2}\right)\ket{g,0}-\Lambda\ket{e,1}+\Lambda^2 \sqrt{2} \ket{g,2},
\end{equation}
where $\Lambda=g/(2\omega)$.

\section{\label{sec: An example of transfer protocol}An example of transfer protocol}

Here, we provide more details on the transfer protocol briefly described at the end of Sec.~\ref{sec: Rabi model}, concerning the Rabi model in the case when the environment is at zero temperature.
First, we suppose that system $A$, which we previously treated as a two-level system, is a three-level system the intermediate third level $\ket{u}$ of which did not participate in the interaction with system $B$ during the thermalization process.
Then, the Hamiltonian of system $A$ has to be written as follows:
\begin{equation}
\label{eq: third level}
H_A=\frac{\hbar \omega}{2}\dyad{e}-\hbar \gamma \dyad{u} - \frac{\hbar \omega}{2} \dyad{g},
\end{equation}
where $\abs{\gamma}<\omega/2$.
We also suppose to have at our disposal a number $N_c$ of systems $D_i$, with the same spectrum of system $A$, in the ground state $\ket{g}$ (these copies are free resources because they are, initially, in the thermal state at $T=0$).

The main idea is to use system $A$ and systems $D_i$ to charge system $C$ through interactions modeled with the Jaynes-Cummings Hamiltonian.
The external harmonic oscillator is chosen to be resonant with the transition connecting states $\ket{e}$ and $\ket{u}$.
The interaction with $C$ will be assumed to involve only these states.
By doing this and taking $\abs{\gamma}<\omega/2$, we assure that $\ket{u}$ is never the ground state in each part of the thermalization protocol and that the interactions with system $C$ are one-way energy transfers from systems $A$ and $D_i$ to system $C$.

The $N_c$ systems $D_i$ interact with system $B$ through the Jaynes-Cummings Hamiltonian:
\begin{equation}
\label{eq: three-level system B interaction}
H_{JC}= \hbar g_B \prt{a \sigma_+ + \ad \sigma_-}.
\end{equation}
To analyze the simplest situation, each copy interacts with the harmonic oscillator for the same time $t_B$.
Turning on this interaction does not require energy because the initial state of the three-level systems is the ground state and no energy is required also for turning off the interaction because $H_{JC}$ commutes with the total Hamiltonian.
By suitably choosing the time $t_B$, system $B$ will be nearly depleted and the energy will be stored in the three-level systems.

Using Eqs.~\eqref{eq: the ground state}, \eqref{eq: form of ground state} and \eqref{eq: parity of coefficients}, the reduced density matrix of system $B$ after the thermalization protocol can be rewritten as follows:
\begin{equation}
\label{eq: no odd coherences}
\rho_B^\rth = \sum_{n,m}\frac{c_n c_m}{2}\prtq{1+(-1)^{n+m}}\dyad{n}{m},
\end{equation}
where we have used the fact that the coefficients $c_n$ are real.
Under the Jaynes-Cummings evolution, a two-level system and a harmonic oscillator in the state $\ket{n+1}$ undergo the following transformation \cite{Haroche_book}:
\begin{equation}\label{JCevo1}
	\ket{g,n+1} \rightarrow -i \sin \alpha_n\ket{e,n} + \cos \alpha_n\ket{g,n+1},
\end{equation}
where $\alpha_n = g_B t_B \sqrt{n+1}$.
A simple calculation shows that, after the interaction, the three-level system is in a mixed state without coherences:
\begin{equations}
	\rho_{D_i} &= \sum_n c_n^2 \prtb{\sin^2 \alpha_{n-1}\dyad{e} + \cos^2 \alpha_{n-1}\dyad{g}}.
\end{equations}
Moreover, the new state of the cavity is of the same form of Eq.~\eqref{eq: no odd coherences}, therefore none of the three-level systems acquires coherences in the energy basis.

\begin{figure}[t!]
	\centering
	\includegraphics[width=0.48\textwidth]{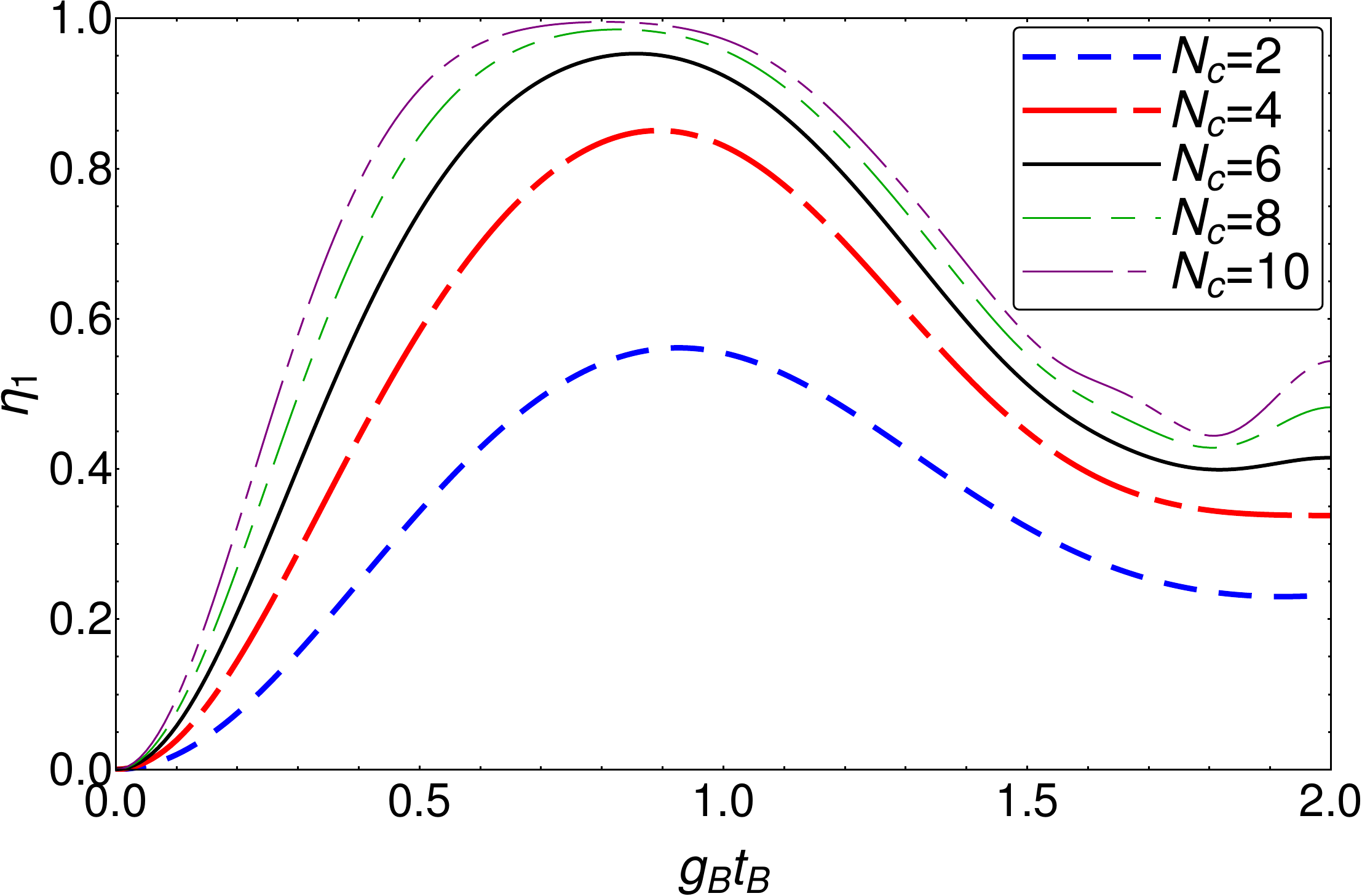}
	\caption{Graph of the efficiency ($\eta_1$) of the extraction part of the transfer protocol as a function of the interaction time $t_B$ for different values of $N_c$.
	}
	\label{fig:grafico_efficienza_estrazione}
\end{figure}

In Fig.~\ref{fig:grafico_efficienza_estrazione}, we plot the efficiency of this energy transfer against the interaction time $t_B$, for $g=1.6 \ \omega$ and different values of $N_c$.
This efficiency is defined as the ratio between the energy acquired by $D_i$ and the energy, $W_B$, that was stored in system $B$, i.e., $\eta_1=\prt{\hbar \omega \sum_{i=1}^{N_c} p_i}/W_B$.
Fig. \ref{fig:grafico_efficienza_estrazione} shows that a great part of $W_B$ ($W_B\simeq 2.49\ \hbar\omega$) can be extracted in this way by properly choosing the interaction time $t_B$.
For comparison, for a Rabi oscillation we have  $g_B t_B = 2 \pi$.
This figure also suggests considering $N_c=8$ in view of the fact that increasing this number raises the extracted energy by a very small amount.
Moreover, the quantity of energy extracted in this way seems to be robust to little variations of $t_B$.
We observe that in general the efficiency $\eta_1$ could be raised (and/or smaller values of $N_c$ could be used) by choosing different interaction times for each of the three-level systems.
The remaining energy of system $B$ will be dissipated in the thermal bath.

In order to charge system $C$, we have then $N_c+1$ three-level systems (the system $A$ and the $N_c$ systems $D_i$) with different excited populations.
The system $A$ and the systems $D_i$ interact with system $C$ through:
\begin{equation}
\label{eq: transfering Hamiltonian}
H'_{JC}= \hbar g_C \prt{c \sigma'_+ + c^\dagger \sigma'_-},
\end{equation}
where $c(c^\dagger)$ is the annihilation (creation) operator for $C$, which has Hamiltonian $H_C= \hbar\omega' \hat{n}_c$, with $\hat{n}_c=c^\dagger c$ and $\omega'=\omega/2+\gamma$.
We recall that this interaction with a system $D_i$ can take place while another three-level system interacts with system $B$.

For every interaction with system $C$, the initial state of any $D_i$ is of the kind
\begin{equation}
	p_e \dyad{e} + 0 \dyad{u} + p_g \dyad{g}.
\end{equation}
To compute the Jaynes-Cummings evolution under a time $t_C$, we can make use of the following transformation, concerning a two-level system in the excited state $\ket{e}$ and an harmonic oscillator in the state $\ket{n}$  \cite{Haroche_book}:
\begin{eqnarray}
& \ket{e,n}   \rightarrow  \cos \prt{g_C t_C \sqrt{n+1}}\ket{e,n}  \nonumber \\
& \qquad -i\sin \prt{g_C t_C \sqrt{n+1}}\ket{u,n+1}.
\end{eqnarray}
It is then easy  to show that each number state of the harmonic oscillator transforms as follows:
\begin{equation}
	\dyad{n} \rightarrow \gamma_n\dyad{n}+ (1-\gamma_n)\dyad{n+1},
\end{equation}
where $\gamma_n = p_e \cos^2 \prt{g_C t_C \sqrt{n+1}} + p_g$.
Then, if the initial state of system $C$ is a state with no coherences in its energy basis, it will never gain coherences from this interaction.
In this case, the initial state of system $C$ is the ground state $\dyad{0}$.

\begin{figure}
	\centering
	\includegraphics[width=0.48\textwidth]{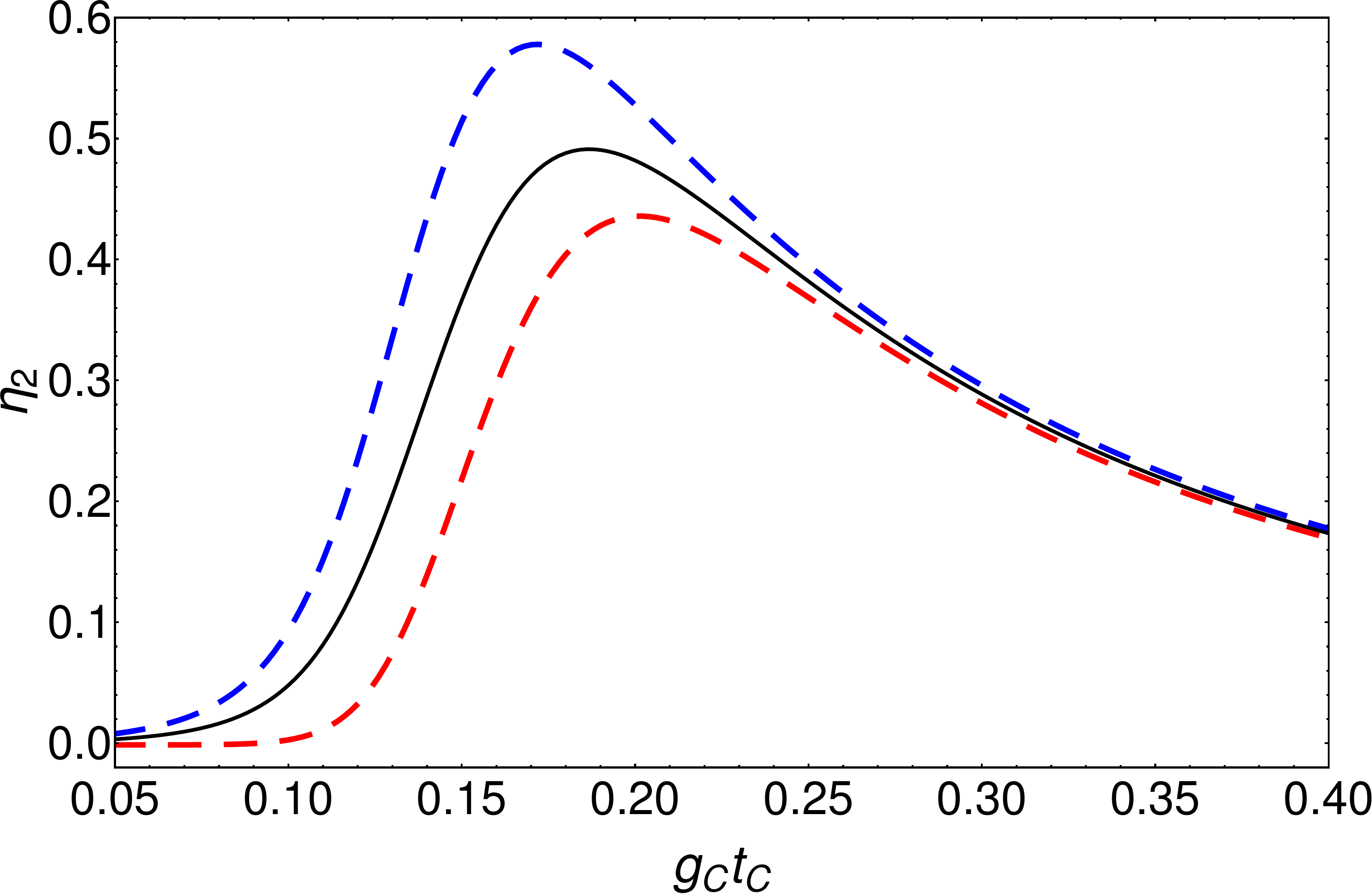}
	\caption{Graph of the efficiency ($\eta_2$) of the charging part of the transfer protocol, for $N=100$ iterations, as a function of the interaction time $t_C$ for $N_c=8$, $g_B t_B=0.84 $ and $\gamma=0.4\ \omega$, which implies $\omega'=0.9\ \omega$.	
	The solid line represents the efficiency obtained using the average energy, while the dotted lines represent the efficiencies obtained using the average energy plus or minus the standard deviation.
	}
	\label{fig:grafico_efficienza_energiaC_N=8}
\end{figure}

\begin{table}
	\begin{tabular}{ ||c|c|| }
		\hline
		\multicolumn{2}{|c|}{Parameters}\\
		\hline
		$W_A= p_a\, \hbar \omega \simeq 0.44\ \hbar\omega$ & $W_B\simeq 2.49\ \hbar\omega$\\
		\hline
		$N=100$ & $N_c=8$\\
		\hline
		$g=1.6\ \omega$ & $\gamma=0.4\ \omega$\\
		\hline
		\multicolumn{2}{|c|}{Optimal interaction times and results}\\
		\hline
		$g_B t_B=0.84 $ & $g_C t_C=0.26 $\\
		\hline
		$\eta_1\simeq 0.985$  & $\eta_2\simeq0.361$  \\
		\hline
		$\eta_T \simeq 0.357$ & $E_C\simeq\prt{104.60 \pm 3.29} \hbar\omega$\\
		\hline
	\end{tabular}
	\caption{All relevant values of the transfer protocol for a specific choice of the parameters.}
	\label{table}
\end{table}

Again, to analyze the simplest situation, each three-level system will interact with the harmonic oscillator for the same time $t_C$.
Cycle after cycle, each three-level system meets system $C$ in a different state, in general, so that the efficiency of this part of the whole protocol depends on the number of cycles.
After having interacted with system $C$, the three-level systems are reinitialized through thermalization and ready to start another cycle of the global protocol.
The average efficiency per cycle of this energy transfer to system $C$ is equal to the ratio of the energy stored in it after $N$ cycles divided by $N$ and the total transferable energy of the three-level systems before the interaction:
\begin{equation}\label{eff2}
\eta_2 = \frac{E_C/N}{\hbar \omega\prt{\sum_{i=1}^{N_c}p_i + p_a}}
=\frac{\omega'}{\omega}\frac{\ev{\hat{n}_C}/N}{\prt{\sum_{i=1}^{N_c}p_i + p_a}},
\end{equation}
where $p_a$ is the excited population of system $A$ at the end of the thermalization protocol (see Eq.~\eqref{A state after therm} for our specific model).

In Fig.~\ref{fig:grafico_efficienza_energiaC_N=8}, we plot the efficiency of this part of the transfer protocol using the three-level systems of the previous part and $100$ iterations of the whole process as a function of $t_C$.
The plot also shows the behavior of the standard deviation.
As one can see, a maximum efficiency of the order of $50\%$ can be achieved for $g_C t_C\simeq 0.18$.
However, by choosing a larger value for $t_C$ we can obtain smaller values for the standard deviation, thus improving the analogy between system $C$ and an ordinary battery, since $C$ is in a mixed state with a relatively high energy and small standard deviation, e.g. $E_C\simeq \prt{104.60 \pm 3.29} \hbar\omega$ for $g_C t_C=0.26 $.

We think that $N=100$ is a suitable compromise to show the iterability of the process while keeping reasonable (at least in principle) the assumption that  the dissipation of system $C $ is negligible.
We have also considered other values of $N$ (for example $N=25$ and $1000$), observing that the results for the efficiency and the standard deviation do not change qualitatively.
We finally observe, from Fig.~\ref{fig:grafico_efficienza_energiaC_N=8}, that the optimal interaction time is much lower than the time of a Rabi oscillation $(g_C t_C =2 \pi$) and that the protocol is robust to small variations of $t_C$.

An interesting feature emerging from numerical simulations is that if we vary the populations $p_i$ by keeping fixed their sum, the plot in Fig.~\ref{fig:grafico_efficienza_energiaC_N=8} almost does not change, for $N$ sufficiently high.
This means that this result is solid with respect to the number of copies $N_c$ and to variations of the populations.

The total efficiency of the complete transfer protocol can be calculated as follows:
\begin{equations}
	\label{eq: total efficiency transfer protocol}
	\eta_T=\frac{W_A+\eta_1 W_B}{W_A+W_B}\times \eta_2.
\end{equations}
In table \ref{table} we report the efficiencies $\eta_1$, $\eta_2$ and $\eta_T$ for some specific values of the relevant parameters, as well as the optimal interaction times (up to the second decimal digit) found through numerical simulations.

One can also estimate the minimum amount of time a cycle of the complete protocol takes when considering both the thermalization  and the transfer protocol.
	By neglecting the $A$-$B$ interaction switches, we get
	\begin{equation}
	T= (t_3 - t_2) + N_c \max \prtg{t_B,t_C}.
	\end{equation}
If $t_B > t_C$, this result is obtained by considering that when the last three-level system of a cycle ends its interaction with system $B$, system $A$ can already be ready to start another thermalization process, i.e., the next cycle of the complete protocol.
If $t_C > t_B$, the last of the copies has to wait $N_c$ interactions of other three-level systems with system $C$.
When its interaction begins, system $A$ can already start the thermalization protocol.
It follows that for $N_c$ and $N$ not too large, depending on the actual physical implementation, the dissipative processes of system $C$ during its charge could be effectively negligible, as we assumed here.

\end{document}